\documentclass[achemso,reprint,english]{revtex4-1}
\usepackage[T1]{fontenc}
\usepackage{color}
\usepackage{amsmath}
\usepackage{amssymb}
\usepackage{graphicx}
\usepackage{esint}
\usepackage{lineno}
\usepackage{tabularx}

\makeatletter
\@ifundefined{textcolor}{}
{%
 \definecolor{BLACK}{gray}{0}
 \definecolor{WHITE}{gray}{1}
 \definecolor{RED}{rgb}{1,0,0}
 \definecolor{GREEN}{rgb}{0,1,0}
 \definecolor{BLUE}{rgb}{0,0,1}
 \definecolor{CYAN}{cmyk}{1,0,0,0}
 \definecolor{MAGENTA}{cmyk}{0,1,0,0}
 \definecolor{YELLOW}{cmyk}{0,0,1,0}
}

\renewcommand{\fnum@figure}{\textbf{Figure~\thefigure}}
\usepackage{units}
\usepackage{lipsum}

\newcommand{\vct}[1]{{\mathbf{#1}}}
\newcommand{\dgr}{{$^{\circ}$}}

\usepackage{babel}

\makeatother

\usepackage{babel}
\begin{document}

\title{Tunable Phases of Moir\'e Excitons in van der Waals Heterostructures}

\author{Samuel Brem}
\author{Christopher Linder\"alv}
\author{Paul Erhart}
\author{Ermin Malic}
\affiliation{
    Chalmers University of Technology,
    Department of Physics,
    41296 Gothenburg, Sweden
}
\begin{abstract}
Stacking monolayers of transition metal dichalcogenides into a heterostructure with a finite twist-angle gives rise to artificial moir\'e superlattices with a tunable periodicity. As a consequence, excitons experience a periodic potential, which can be exploited to tailor optoelectronic properties of these materials. While recent experimental studies have confirmed twist-angle dependent optical spectra, the microscopic origin of moir\'e exciton resonances has not been fully clarified yet. Here, we combine first principle calculations with the excitonic density matrix formalism to study transitions between different moir\'e exciton phases and their impact on optical properties of the twisted MoSe$_2$/WSe$_2$ heterostructure. At angles smaller than 2\dgr{} we find flat, moir\'e trapped states for inter- \textit{and} intralayer excitons. This moir\'e exciton phase drastically changes into completely delocalized states already at 3\dgr{}. We predict a linear and quadratic twist-angle dependence of excitonic resonances for the moir\'e-trapped and delocalized exciton phase, respectively. Our work provides microscopic insights opening the possibility to tailor moir\'e exciton phases in van der Waals superlattices.

\end{abstract}
\maketitle

\begin{figure}[t!]
 \centering
\includegraphics[width=\columnwidth]{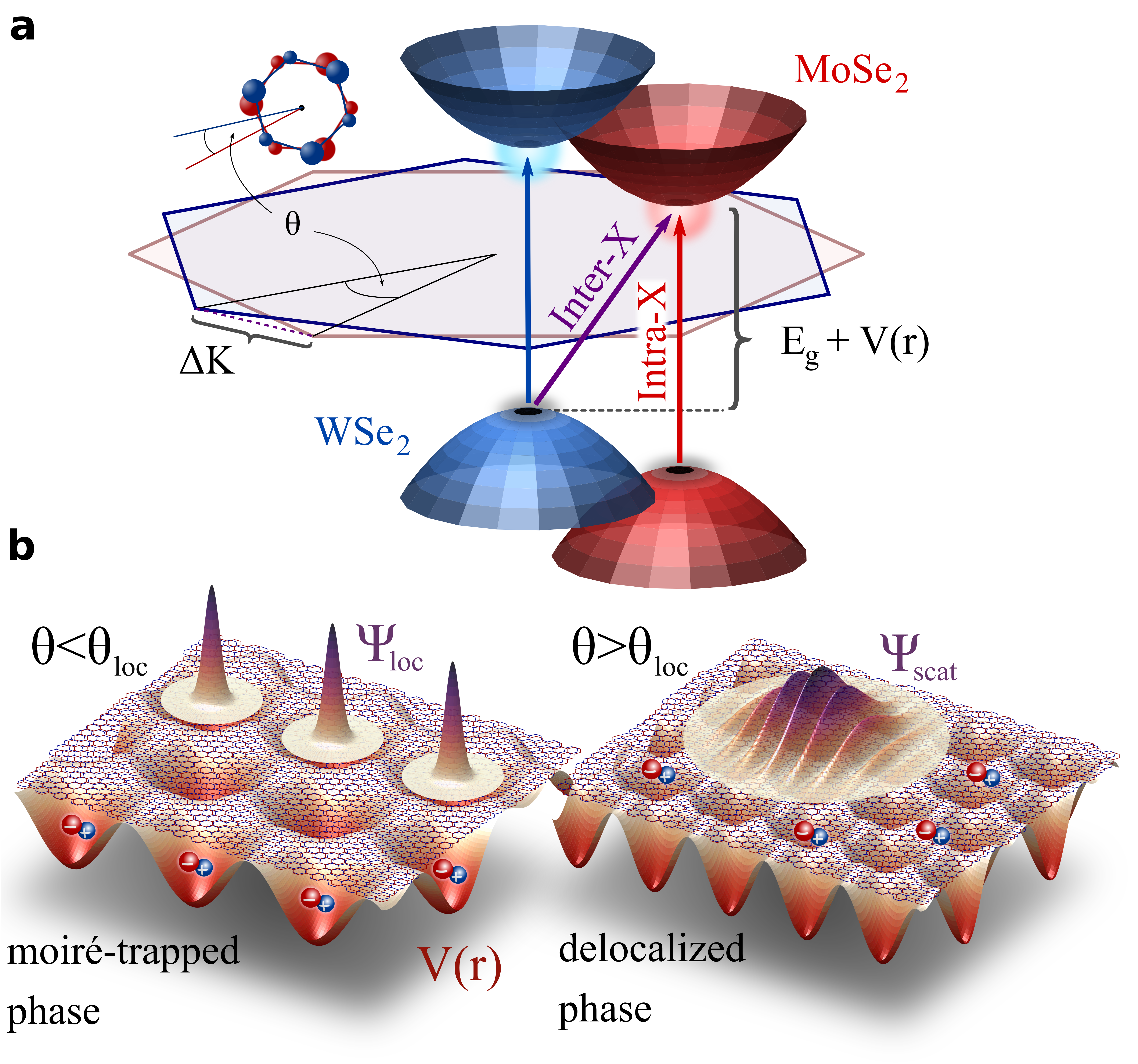}
\caption{ Schematic illustration of moire excitons
    \textbf{a} Van der Waals heterostructures can host intra- (red and blue) as well as interlayer excitons (purple) at the K points (corners) of their hexagonal Brillouin zones.
    A finite stacking angle $\theta$ gives rise to a mismatch of the dispersions in momentum space and at the same time leads to a spatially periodic moir\'e potential $V(r)$.
    \textbf{b} Depending on the length of the moire period, the exciton center-of-mass motion can become quantized, leading to either moir\'e-trapped (left) or delocalized scattering states (right).
}
\label{fig:scheme} 
\end{figure}

Atomically-thin quantum materials, such as graphene and transition metal dichalcogenides (TMDs), obtain their unique properties by hosting strongly interacting quasi-particles.
In particular, the strong Coulomb interaction in TMDs allows to study and control the dynamics of excitons \cite{he2014tightly, chernikov2014exciton, wang2018colloquium,mueller2018exciton}, which are atom-like quasi-particles composed of Coulomb-bound electrons and holes.
In a TMD monolayer, excitonic properties can be widely tuned through strain \cite{niehues2018strain,aslan2018strain} and dielectric engineering \cite{latini2015excitons,steinleitner2018dielectric, raja2019dielectric}.
Moreover, two monolayer TMDs can be vertically stacked to form a type-II heterostructure giving rise to spatially indirect and therefore ultra-stable interlayer excitons \cite{rivera2015observation,miller2017long, merkl2019ultrafast}, as depicted in Figure \ref{fig:scheme}a.
Recent studies \cite{jin2019observation,tran2019evidence,seyler2019signatures,alexeev2019resonantly} have shown that artificial moir\'e superlattices can be created by vertically stacking monolayers with a finite twist-angle, giving rise to a tunable modification of exciton features in optical spectra.
Theoretical studies on homo- \cite{wang2017interlayer} and resonantly aligned heterobilayers \cite{ruiz2019interlayer} show that the twist-tunability can be attributed to a stacking-dependent hybridization of intra- and interlayer excitons \cite{merkl2020twist,kiemle2020control,brem2020hybridized}.
Moreover, recent studies on heterostructures with a large band-offset, where hybridization at the K point is negligible, suggest that moir\'e patterns lead to a spatially varying interlayer distance and electronic bandgap \cite{zhang2017interlayer, wu2017topological, wu2018theory}.
Consequently, excitons experience a moir\'e periodic potential, which can be exploited to tailor exciton transport properties \cite{wu2018hubbard} or even create tunable quantum emitter arrays \cite{yu2017moire}. 

To be able to exploit the technological potential and to design future experiments on van der Waals stacked superlattices, microscopic insights into moir\'e excitons are needed. First-principle computations can provide important insights about the electronic structure in these systems \cite{liu2014evolution,naik2018ultraflatbands,lu2019modulated, guo2020shedding}, but they are fundamentally limited to perfectly aligned or very specific twisted structures with relatively small supercells.
However, most exciting phenomena, such as superconductivity in magic-angle bilayer graphene \cite{cao2018unconventional, bistritzer2011moire}, are expected to occur at very small twist-angles with large supercells including thousands of atoms. Therefore, effective continuum models \cite{wu2017topological, wu2018theory} as used for bilayer graphene are needed to describe moir\'e excitons at small twist-angles.

In this work we combine first-principle calculations with the excitonic density matrix formalism to develop a material-specific and realistic exciton model for small-angle twisted MoSe$_2$/WSe$_2$ heterostructures.
Based on a microscopic approach we calculate the band structure and wave functions of intra- and interlayer excitons within a twist-tunable moir\'e lattice as well as the resulting optical response of these compound particles. For a range of small twist angles, we predict completely flat exciton bands for both intra- as well as interlayer excitons corresponding to moir\'e trapped, localized quantum emitters.
However, we reveal that this moir\'e exciton phase dramatically changes already at a twist-angle of 3\dgr{} leading to completely delocalized wave functions, as depicted in Figure \ref{fig:scheme}b. We find the emergence of multiple moir\'e exciton peaks in the absorption, whose spectral shifts with varying twist-angle are characteristic for the trapped or delocalized phase.
While previous studies predicted rather shallow moire potentials for intralayer excitons, we find that GW corrections of the electronic band structure lead to much larger intralayer band gap variations and a moir\'e-induced double peak structure characterizing intralayer excitons. Furthermore, we also show that excitonic effects lead to a reduction of the effective moire potential depth with increasing twist-angles. Overall, our work provides microscopic insights and a comprehensive picture that unifies different moir\'e exciton phases and their optical signatures.    


\section*{Results}
\textbf{Moir\'e Potential.}
In this work, we study the type-II MoSe$_2$/WSe$_2$ heterostructure, focusing on the energetically lowest interband transitions at the K points of the hexagonal Brillouin zone.
First-principle calculations of the electronic structure in this material suggest that the wave functions of states at these band edges are strongly layer-polarized \cite{gillen2018interlayer, lu2019modulated}, i.e. the quasi-planar monolayer eigenstates only weakly hybridize in this region of the band structure.
However, recent experimental as well computational studies suggest that the valence and conduction band energies at the K point nonetheless significantly vary with the geometrical alignment of the two layers \cite{wu2017topological,zhang2017interlayer,yu2017moire}.
Consequently, in a twisted stacking configuration, where the local alignment of atoms is periodically changing, a spatially varying moir\'e potential is expected to emerge.

We model the moir\'e Hamilton operator by assuming that the energy variation $V$ results from interactions of the d-orbitals (composing electronic states at the K point) with the effective atomic potentials of the neighboring layer (see Supplementary):
\begin{equation}
\label{eq:Moir\'ePot}
H = \sum_{\vct{k,q}} V_\vct{q} a^\dagger_{\vct{k+q}} a_{\vct{k}}+\text{h.c.},\hspace{0.25cm} V_\vct{q}= v_0 \sum^2_{n=0} e^{ i \hat{\vct{G}}_n \vct{D}} \delta_{\vct{q},\vct{g}_n}.
\end{equation}
Here, $a_{\vct{k}}$ is the field operator of an electron with momentum $K+\vct{k}$, which interacts with the moir\'e potential determined by the lateral displacement of the two layers $\vct{D}$ and the fundamental translation $\vct{G}_{l}$ of the reciprocal lattice of layer $l$ with $\hat{\vct{G}}_n=C_3^n(\vct{G}_1+\vct{G}_2)/2$ and $\vct{g}_n=C_3^n(\vct{G}_{2}-\vct{G}_1)$.
The operator $C_3$ rotates a vector by 120\dgr{}, reflecting the hexagonal symmetry of the two subsystems. The form of Eq. \eqref{eq:Moir\'ePot} is equivalent to phenomenological formulas applied in previous studies \cite{wu2017topological,yu2017moire, wu2018theory}, but it is here derived  from a microscopic approach describing the modification of electronic energies due to the van der Waals potential of the adjacent layer.
When neglecting the small difference in lattice constants between MoSe$_2$ and WSe$_2$ and setting the twist angle to zero, Eq.~\eqref{eq:Moir\'ePot} collapses to a simple, spatially independent energy shift resulting from the presence of the other layer. 
This shift, however, depends on the lateral displacement $\vct{D}$ that determines the atomic registry between the layers.
Hence, the complex-valued parameter $v_0$ can be obtained by fitting the band- and layer-dependent potential in Eq.~\eqref{eq:Moir\'ePot} to the band energies obtained by first-principle methods for different high-symmetry stackings.  

\begin{figure}[t!]
\centering
\includegraphics[width=\columnwidth]{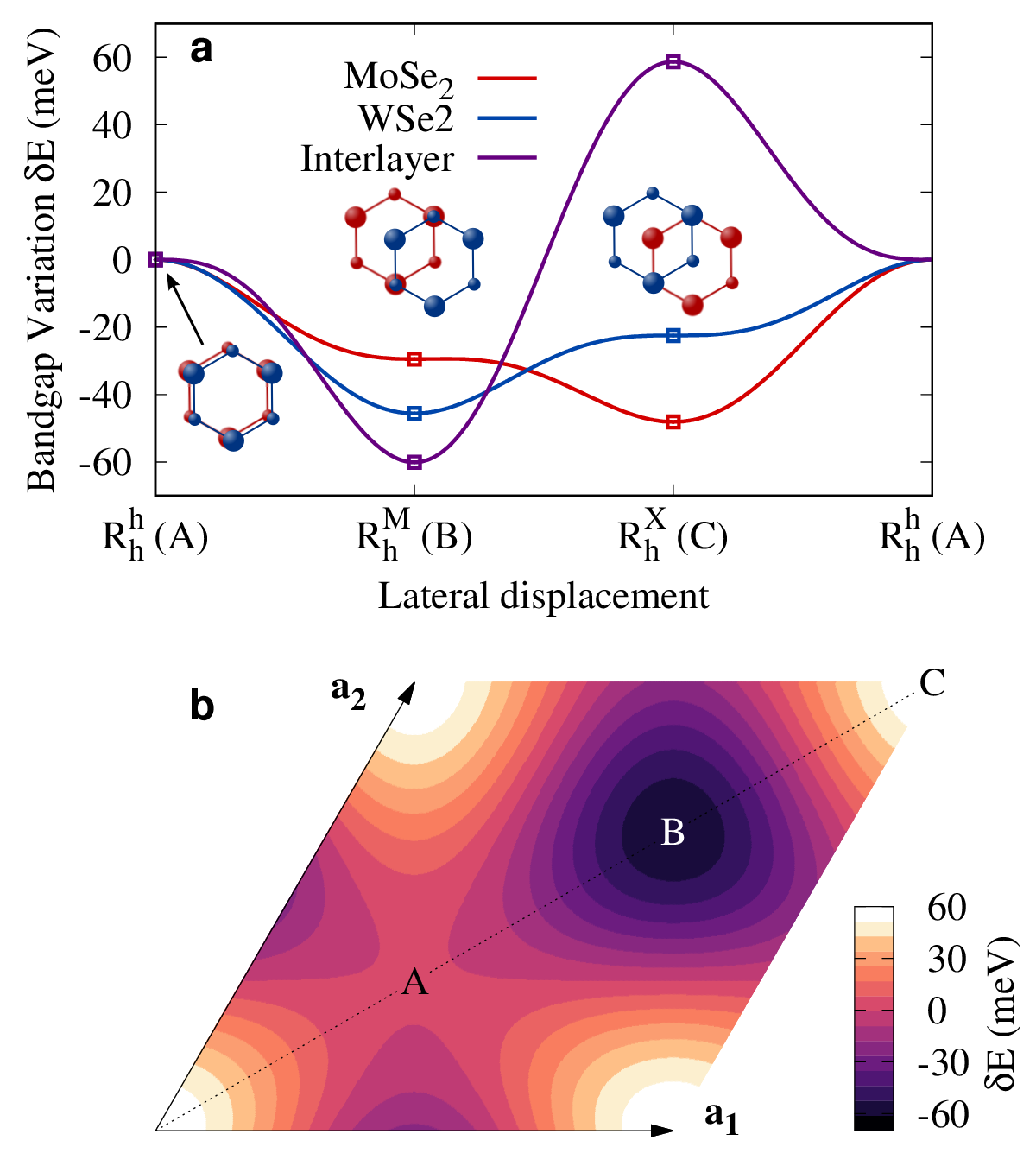}
\caption{
    Moir\'e potential in the MoSe$_2$/WSe$_2$ heterostructure.
    \textbf{a} Bandgap variation for the two intralayer transitions (red and blue) and the energetically lowest interlayer transition (purple) at the K point as a function of a lateral shift between the layers at zero twist-angle.
    The dots show the energies calculated using DFT+ G$_0$W$_0$, while the lines are extracted from a fitted microscopic model.
    \textbf{b} Interlayer bandgap variation in a twisted heterostructure, where $\vct{a}_i$ are the two fundamental translations of the moir\'e superlattice.
    A, B and C denote high-symmetry sites in the moir\'e pattern, corresponding to local atomic registries specified in \textbf{a}.
}
\label{fig:potential} 
\end{figure}

Figure \ref{fig:potential}a shows the bandgap variations for different stackings obtained from first-principles calculations (dots, see Supplementary) as well as the fitted microscopic model (lines).
The obtained energy shifts agree well with previously reported values for DFT+GW calculations \cite{lu2019modulated} and the analytical model excellently reproduces the displacement dependence by fixing only two parameters per band. 
The advantage of the developed microscopic model is that we can now use an analytical expression to model the moir\'e potential in an arbitrarily twisted material.
Figure \ref{fig:potential}b shows the predicted interlayer bandgap variation as a function of the spatial coordinate within a moir\'e supercell, where $\vct{a}_1$ and $\vct{a}_2$ are fundamental translations of the superlattice.
We find a triangular symmetry with a total minimum at the point denoted with B (local R$^M_h$ stacking), a total maximum at C (R$^X_h$) and an intermediate inflection point A (R$^X_h$).
Based on the derived analytical model, we can now make a material-specific study of exciton characteristics in the moir\'e potential of a twisted heterostructure. 
\\

\textbf{Excitonic Moir\'e Mini-Bands.}
Due to the strong Coulomb interaction in 2D systems, the band-edge excitations in TMDs are governed by excitons \cite{wang2018colloquium}.
The total energy of an exciton is given by the band gap and the Coulomb binding energy.
Consequently, a spatial variation of the bandgap throughout the moir\'e pattern creates an effective potential influencing the exciton center-of-mass (CoM) motion.
To account for the excitonic character of the elementary excitations in TMDs, we transform the Hamiltonian for conduction ($c$) and valence band electrons ($v$) into an exciton basis \cite{haug1984electron,katsch2018theory} $X^\dagger_\vct{Q}=\sum_\vct{k} \psi^\ast_\vct{k} c^{\dagger}_{\vct{k+\alpha Q}} v_{\vct{k-\beta Q}}$ with the ground state exciton wave function $\psi_\vct{k}$, CoM momentum $\vct{Q}$ and mass coefficients $\alpha=m_e/(m_e+m_h)$, $\beta=1-\alpha$.
We thereby obtain an effective single-particle Hamiltonian
\begin{eqnarray} \label{eq:H_X}
H &=& \sum_{\mu\vct{Q}} \mathcal{E}^\mu_{\vct{Q}} X^\dagger_{\mu\vct{Q}} X_{\mu\vct{Q}} + \sum_{\mu\vct{Q,q}} \mathcal{M}^\mu_\vct{q} X^\dagger_{\mu\vct{Q+q}} X_{\mu\vct{Q}},
\end{eqnarray}
containing the interaction-free part with the exciton CoM dispersion $\mathcal{E}_{\mu\vct{Q}}$ and the effective exciton moir\'e potential incorporating the spatial bandgap variation [Eq.~\eqref{eq:Moir\'ePot}] convoluted with the exciton wave function (cf. Supplementary).
The index $\mu$ is referring to different exciton states, either composed of electrons and holes within the same (intralayer exciton) or different layers (interlayer exciton).
Equation~\eqref{eq:H_X} describes excitons moving in the effective periodic potential of the moir\'e superlattice, analogous to electrons in a crystal.
Hence, the spectrum of its eigenstates is given by a series of subbands $\zeta$ defined in the mini-Brillouin zone (mBZ) spanned by the reciprocal lattice vectors of the moir\'e pattern $\vct{g}_n$.
We therefore diagonalize Eq.~\eqref{eq:H_X} for each exciton species via a zone-folding approach yielding
\begin{eqnarray} \label{eq:H_diag}
H &=& \sum_{\zeta \vct{Q}} E^\zeta_{\vct{Q}} Y^\dagger_{\zeta\vct{Q}} Y_{\zeta\vct{Q}},\hspace{0.25cm}Y_{\zeta\vct{Q}}=\sum_{i,j} \mathcal{C}^\zeta_{ij\vct{Q}} X_{\vct{Q}+i\vct{g}_0+j\vct{g}_1}. 
\end{eqnarray}
In this framework, the emergence of moir\'e excitons $Y$, analogous to Bloch waves of electrons in a crystal, can be understood as periodic superposition of CoM momenta, which has important consequences for optical selection rules of moir\'e excitons \cite{wu2017topological}. 

\begin{figure}
\centering
\includegraphics[width=\columnwidth]{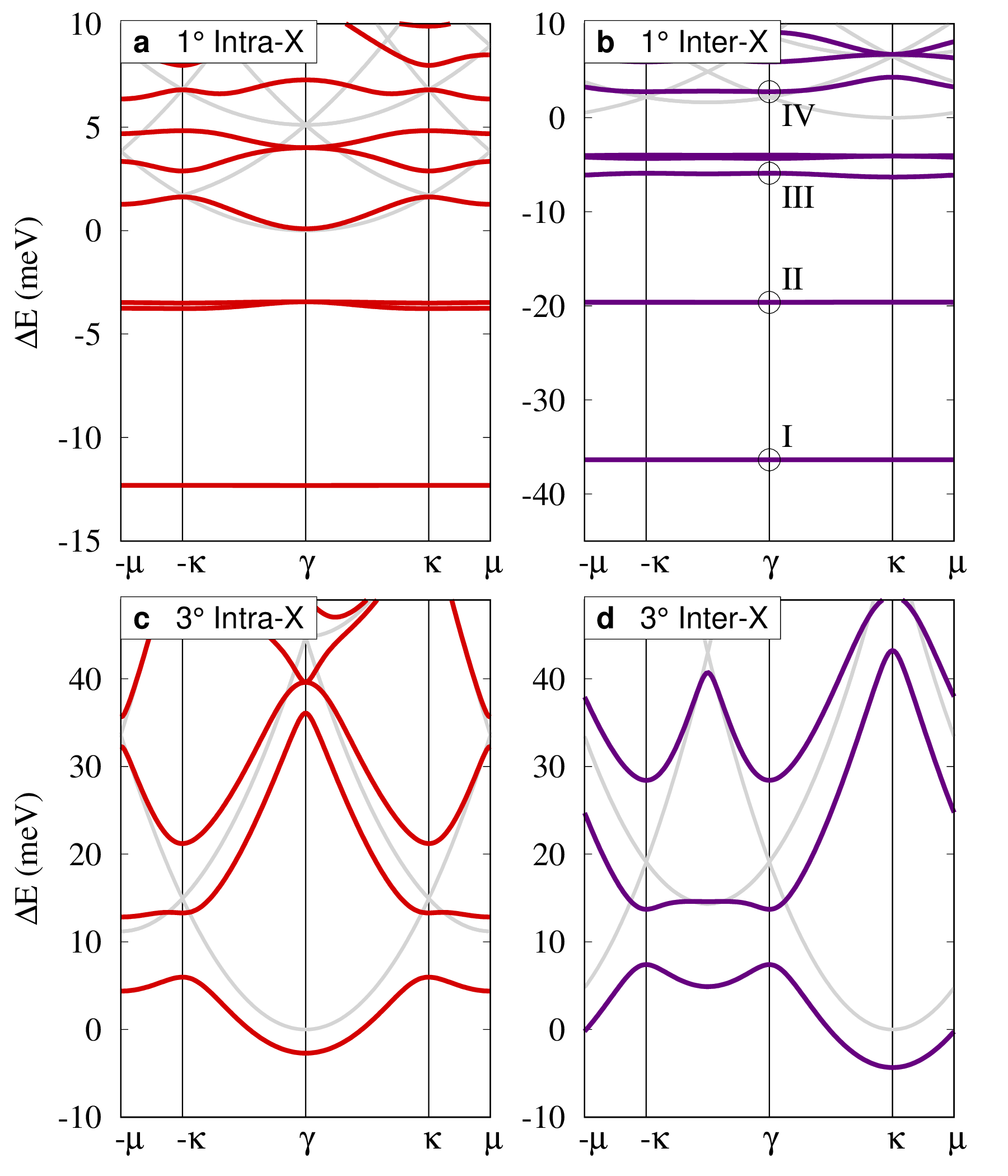}
\caption{
    Moir\'e exciton mini-bands for the twisted MoSe$_2$/WSe$_2$ heterostructure.
    While \textbf{a} and \textbf{b} show the MoSe$_2$ intra- and the lowest interlayer exciton at 1\dgr{}, respectively, \textbf{c} and \textbf{d} illustrate the corresponding bands at 3\dgr{}.
    At 1\dgr{} both inter- \textit{and} intralayer excitons exhibit flat, moir\'e-trapped states, but only the deeper interlayer potential shows a whole series of localized states (denoted by I-III).
    For larger twist-angles, one finds almost parabolic bands close to a zone-folded free exciton dispersion (gray curves).
    Here, the free particle dispersion is only effectively modified at intersection points, where a strong mixing of different CoM momenta leads to avoided crossings that can be interpreted as standing waves at different sites of the moir\'e potential.
}
\label{fig:BS} 
\end{figure}

The calculated excitonic band structure for a twisted MoSe$_2$/WSe$_2$ bilayer is shown in Fig. \ref{fig:BS} exemplifying the MoSe$_2$ intralayer exciton and the lower interlayer exciton.
To better illustrate the sensitive dependence on the twist-angle, we show the resulting mini-bands for 1\dgr{} and 3\dgr{}.
To this end, the dispersions are plotted along a high-symmetry path of the hexagonal mBZ with the center $\gamma$ and the edges $\kappa=\Delta K$, defined by the mismatch of the two monolayer K points.
In addition to the solution of Eq.~\eqref{eq:H_diag}, we also show the zone-folded free exciton dispersion, corresponding to energies without moir\'e potential (gray lines).
At 1\dgr{}, we find completely flat exciton bands for both inter- \textit{and} intralayer excitons. While flat bands for interlayer excitons have been already predicted in previous studies \cite{wu2018theory}, intraband excitons were believed to have too shallow moire potentials to trap excitons. In this work, we find that GW corrections of the electronic band structure lead to significantly larger intralayer band gap variations in the range of 50 meV (Fig. \ref{fig:potential}) allowing for exciton localization at small angles.
The appearance of non-dispersed bands means that the corresponding exciton states have a vanishing group velocity and that hopping between neighboring moir\'e supercells is completely suppressed.
Therefore, these states can be considered as moir\'e-localized zero-dimensional states analogous to quantum dots.
Thereby the deeper interlayer exciton potential (Fig. \ref{fig:potential}\textbf{a}) gives rise to several flat bands with large separations, while the shallower intralayer potential only allows one localized state and all excited states are spread over the whole crystal.
These higher-order moir\'e bands with a dispersive character can be interpreted as scattering states, i.e. delocalized, free exciton waves with moir\'e-periodic amplitude modulations. 

At 3\dgr{}, all subbands for both exciton species are delocalized, exhibiting an almost parabolic dispersion close to the free particle situation (gray lines).
This results from the shrinking of the moir\'e supercell with increasing twist angle.
Consequently, the kinetic energy of the ground state, that is the zero-point energy, increases with the twist angle until the exciton starts to hop between neighboring moir\'e cells.
However, the dispersion at 3\dgr{} still exhibits band gaps at intersection points of the zone-folded free-particle dispersion.
Here, standing waves emerge from Bragg reflection at the moir\'e potential.
The lowest interlayer exciton in Fig.~\ref{fig:BS}\textbf{d} at the $\gamma$ point is a standing wave with large probability at B sites, while the second lowest state has the largest probability at A sites.
The results shown in Fig.\ref{fig:BS} illustrate that the moir\'e exciton phase (including moir\'e localized or delocalized states) and consequently the excitonic transport properties can be widely tuned in TMD heterostructures, since the general character of the excitonic ground state is very sensitive to changes in the twist-angle-dependent moir\'e period.
\\

\begin{figure}
\centering
\includegraphics[width=\columnwidth]{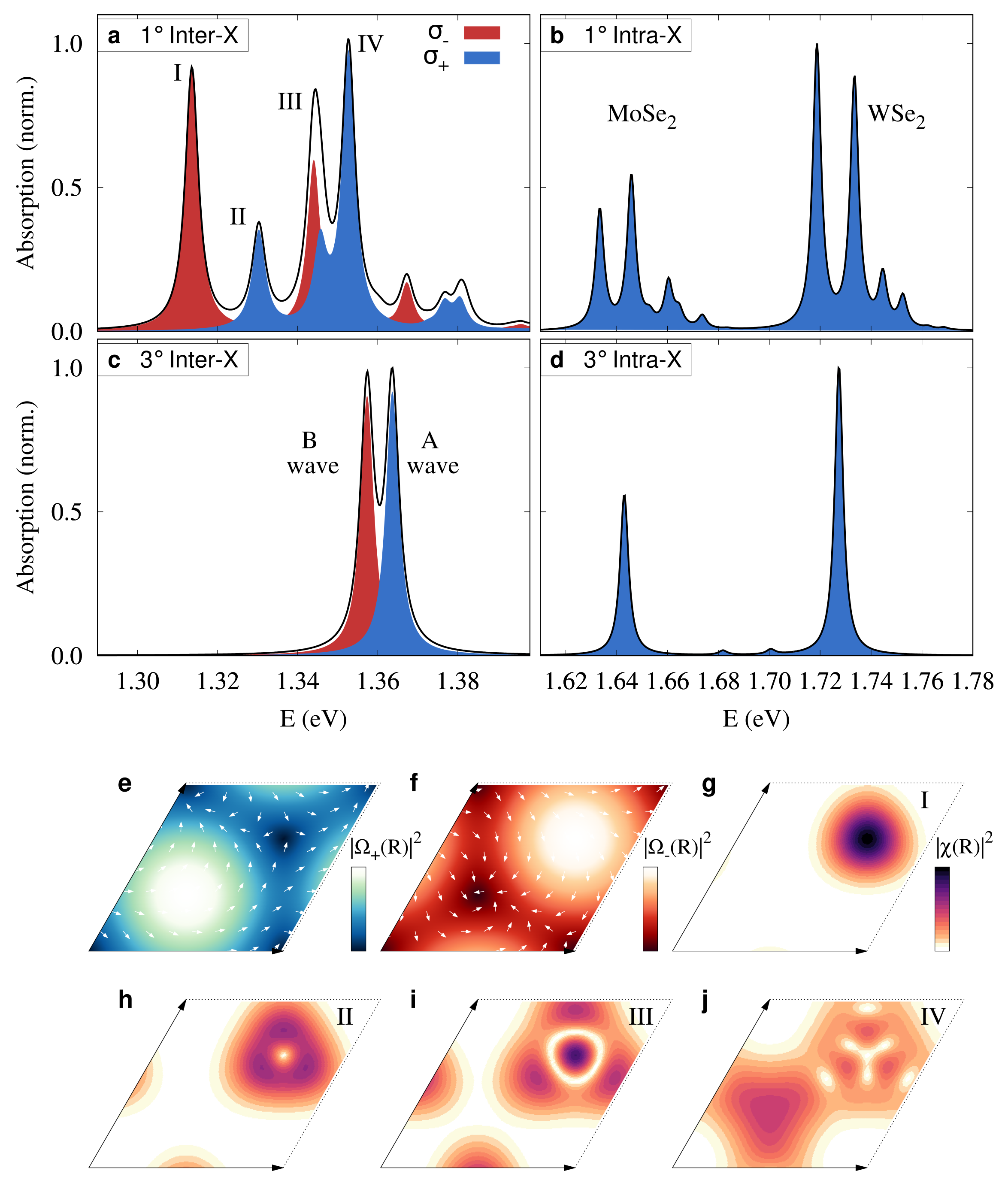}
\caption{
    \textbf{a-d} Moir\'e exciton absorption spectra at 1 and 3\dgr{}.
    The moir\'e-induced mixing of exciton center-of-mass momenta gives rise to a splitting of the single exciton peak into a series of moir\'e resonances at 1\dgr{} for both intra- as well as interlayer excitons. At 3\dgr{} the weak mixing leads to much smaller modifications of the absorption spectrum. While intralayer resonances are fully $\sigma_+$ polarized (blue), interlayer moir\'e excitons show an alternating $\sigma_+$ and $\sigma_-$  polarization.
    \textbf{e} and \textbf{f} show the spatial distribution of the oscillator strength for $\sigma_+$ and $\sigma_-$ polarized light.
    The phase of the dipole matrix element is illustrated as a superimposed vector field.
    \textbf{g}-\textbf{j} Exciton center-of-mass wave functions for the states I--IV denoted in \textbf{a}.
    For s-type states (I, III and IV) the polarization of the corresponding absorption peaks results from the maximum of the wave function, while for p-type states (II) the phase of the matrix element becomes crucial.
}
\label{fig:spectra} 
\end{figure}

\textbf{Moir\'e-Modified Light-Matter Coupling.}
The modulation of the exciton CoM momentum discussed above has a direct impact on the light-matter interaction in twisted heterostructures.
Based on the transformation into a moir\'e exciton basis, we calculate the absorption spectra \cite{kira2006many,koch2006semiconductor} of the twisted MoSe$_2$-WSe$_2$ heterostructure (see Supplementary). Figure \ref{fig:spectra}a-d shows the absorption of the 1 and 3\dgr{} material in the spectral region of the inter- and intralayer excitons. Here, the red (blue) curve corresponds to the absorption coefficient for  left $\sigma_{-}$ (right $\sigma_{+}$) circularly polarized light. 
At 1\dgr{} we find multiple inter- and intralayer exciton peaks, while a perfectly aligned heterostructure would exhibit only one single resonance for each exciton species.
To understand this moir\'e phenomenon, we have to consider the momentum conservation for the light-matter interaction.
In a regular semiconductor, only excitons with zero CoM momentum can be created by or decay into a photon, since the latter has a negligible momentum.
This selection rule is modified in a moir\'e superlattice, where Bragg reflection at the periodic potential can scatter excitons with non-zero momentum into the light cone \cite{yu2015anomalous}.
Consequently, all exciton minibands at the center of the mBZ ($\gamma$ point) can in principle couple to photons.
Their oscillator strength $\Omega_\zeta \propto \mathcal{C}^\zeta_{\vct{Q=0}}$ is determined by their projection onto the original bright state ($Q=0$), i.e. the oscillator strength of the bright exciton is redistributed across all mini-bands according to the momentum spectrum of their wave function and is therefore mostly transferred to the lowest energy states with small kinetic energy.
At 3\dgr{}, we find only a single exciton resonance of the intralayer excitons, resembling the unperturbed $\vct{Q}=0$ exciton.
In contrast, the interlayer exciton still shows two distinct resonances, resulting from the avoided crossing and the two low energy states corresponding to standing waves at A and B sites. 

While all intralayer exciton resonances are fully $\sigma_{+}$ polarized as in the monolayer case, we find alternating polarizations for the observed interlayer peaks.
To explain this behavior microscopically, we show the optical matrix element for the interlayer electron-hole recombination in Fig.~\ref{fig:spectra}\textbf{e} for $\sigma_{+}$ and in Fig. \ref{fig:spectra}\textbf{f} for $\sigma_{-}$ light.
Similar to the spatial variation of the band gap, the locally varying atomic alignment also has an impact on the optical selection rules for electronic transitions across both layers.
While coupling to $\sigma_{+}$ light is strongest at A sites, it vanishes at B sites and the opposite is true for $\sigma_{-}$ light \cite{yu2015anomalous, yu2017moire,wu2018theory}.
Along the matrix elements we also show the exciton CoM wave function for states I--IV in Figure \ref{fig:spectra}\textbf{g-j}.
We find that the states I--III are localized at the potential minimum at the B site, whereas state IV is delocalized and has the largest probability at A sites.
Comparing the overlap between wave functions and the spatial distribution of oscillator strength explains the polarization of the peaks I, III ($\sigma_{-}$) and IV ($\sigma_{+}$).
Interestingly, state II has p-type symmetry and should be dark, because of its angular momentum (phase $\exp{i\phi}$). However, the complex phase of the $\sigma_{+}$ matrix element, shown as the white vector field in Fig.~\ref{fig:spectra}e,  has the opposite winding number (cf. rotation of vector field around B site) so that p-type states are bright in this spatial valley.
Hence, the alternating series of polarizations of the interlayer absorption shown in Fig.~\ref{fig:spectra}a-d results from the ladder of alternating s and p type states and/or probability maxima at A or B sites.
\\

\begin{figure}
\includegraphics[width=0.5\textwidth]{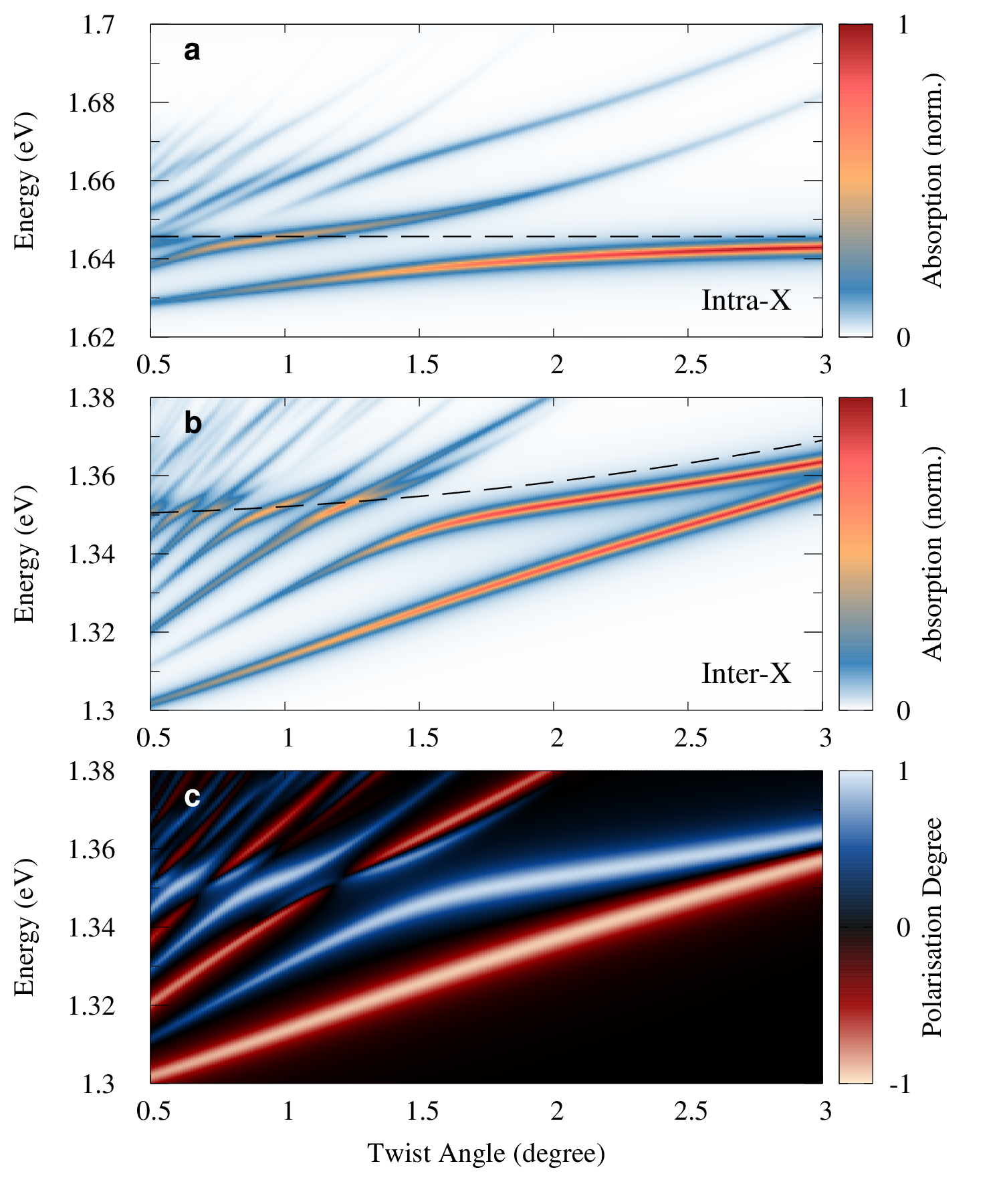}
\caption{
    Absorption spectrum of \textbf{a} intralayer and \textbf{b} interlayer excitons as continuous function of the twist-angle.
    While all signals of the intralayer exciton are circularly polarized, the different interlayer exciton states show an alternating polarization, which is illustrated in \textbf{c}.
    The twist-angle dependence of the multiple moir\'e resonances varies for different exciton phases of moir\'e-trapped and delocalized scattering states.
    For localized states with energies far below the free exciton (dashed lines) resonances shift linearly with the twist angle.
    In contrast, moir\'e resonances stemming from scattering states (energies above or close to the dashed lines) have a quadratic twist-angle dependence reflecting the free-particle dispersion.
}
\label{fig:Abs_scan} 
\end{figure}

\textbf{Twist-tuning Moir\'e Excitonic Phases.}
Now, we investigate how the character and consequently the optical signatures of moir\'e excitons evolve with the twist angle.
Figure \ref{fig:Abs_scan} shows absorption spectra of intra- (a) and interlayer excitons (b) together with their degree of polarization (c) as function of the twist-angle.
In order to understand the intriguing behavior of the appearing moir\'e resonances, we have to distinguish between different moir\'e exciton phases including i) moir\'e-localized states and ii) delocalized scattering states.
For energy levels far below the unperturbed exciton ground states (dashed lines), the energy increases linearly with the twist-angle.
These states correspond to zero-dimensional excitons deeply trapped in a moir\'e potential.
With larger twist-angles, the moir\'e period decreases, which corresponds to a shorter confinement length.
This leads to larger kinetic energies, so that the ground state (zero-point) energy as well as the distance between different localized states increase,  cf. Fig.~\ref{fig:Abs_scan}\textbf{b} for $\theta<1$\dgr{}.
The closer a localized state gets to the free particle edge, the larger its CoM orbital becomes until it starts to overlap with neighboring cells and the exciton becomes delocalized.
For the delocalized moir\'e resonances, we find a quadratic energy dependence on the twist angle.
This can be ascribed to the quadratic dispersion of the scattering states similar to the unperturbed system (cf. Fig.~\ref{fig:BS}\textbf{c} and \textbf{d}).
With larger twist-angles the size of the mBZ increases ($\kappa=\Delta K \propto \theta$) and the intersection point of zone-folded branches with the light cone moves towards higher energies. 

It is important to note that the lowest intralayer exciton resonance stops shifting at a certain twist angle, resembling the unperturbed bright exciton resonance at $\vct{Q}=0$.
In contrast, the interlayer exciton resonance is further shifting upwards in energy.
This does not result from the interaction with the periodic moir\'e lattice, but only reflects the indirect nature of the interlayer exciton in a twisted bilayer, cf. Fig. \ref{fig:scheme}.
Since valence and conduction band are shifted away from each other, the minimum of the CoM dispersion is shifted away from the light cone (compare gray curves in Fig.~\ref{fig:BS}\textbf{b} and \textbf{d}).
Consequently, the bright state is moving up in energy in quadratic fashion, reflecting the dispersion of the free interlayer exciton. 
Finally, we find that the splitting between the two remaining interlayer exciton resonances at angles >2.5\dgr{} (A and B waves resonance) decreases with increasing twist angle.
This splitting is proportional to the effective moir\'e potential acting on the CoM coordinate of the exciton $\vct{R}$, cf. Eq.~\eqref{eq:H_X} reading
$
\mathcal{M}^\mu(\vct{R}) = \langle\mu|V^{c}(\vct{R}+\beta\vct{r})- V^{v}(\vct{R}-\alpha\vct{r})|\mu \rangle.
$
At large moir\'e periods the effective excitonic potential is given by the fluctuations of the band gap $\mathcal{M}\approx V^c-V^v$.
However, when the length scale of the moir\'e period gets in the range of the exciton Bohr radius, the exciton does not interact with the potential like a point particle anymore.
The effective exciton potential is then given by a weighted average of the potential over the space occupied by the exciton.
This explains the decreasing interlayer exciton splitting in Fig.~\ref{fig:Abs_scan} for larger angles, as the exciton starts to occupy larger regions of the moir\'e cell and thereby averages over potential minima and maxima. 


\section*{Discussion}
The presented work provides a consistent microscopic framework to model the properties of moir\'e excitons in twist-tunable Van-der-Waals superlattices. In particular, it allows us to study transitions between different moire exciton phases and their impact on the optical response of different heterostructures.

The fourfold interlayer exciton features shown in Fig. \ref{fig:spectra}a have recently been demonstrated in PL spectra \cite{tran2019evidence}. In particular, the measured peaks at 1\dgr{} exhibited a similar splitting in the range of 20meV and also showed alternating degrees of circular polarization - in excellent agreement with our results.
Moreover, the predicted moir\'e-induced splitting of intralayer excitons was experimentally shown in reflection contrast of WS$_2$/WSe$_2$ \cite{jin2019observation} and photoluminescence in MoSe$_2$/MoS$_2$ \cite{zhang2018moire}. However, the intralayer double-peak structure for MoSe$_2$/WSe$_2$ with a splitting of about 10meV predicted in our work has not been observed yet. Furthermore, there is still no direct evidence for flat moire exciton bands for intra- or interlayer excitons.

While the intralayer exciton features are directly accessible in absorption experiments, spatially indirect interlayer excitons are only visible in PL spectra due to their low oscillator strength. In PL however, other decay channels, such as trions \cite{bellus2015tightly}, defect-bound states \cite{vialla2019tuning} and phonon-assisted sidebands \cite{brem2020phonon}, can become important and might be superimposed on the  moir\'e exciton features. Nevertheless, in a systematic study varying excitation density, temperature and applied electric/magnetic field these possible decay mechanism can be potentially disentangled. Moreover, future progress in stacking techniques might enable deterministic twist-angle studies with small step-sizes of 0.1\dgr{}, which would allow to further experimentally verify the predicted twist-angle-dependent transition from a localized regime (linear shifts) to an unbound/scattering phase of excitons (parabolic shifts). Moreover, the transition from moir\'e-trapped to delocalized states should be evident in exciton diffusion measurements.  

Overall, our work  provides fundamental microscopic insights into exciton localization and light-matter coupling in twisted van der Waals heterostructures and will guide future experimental as well as theoretical studies in this growing field of research. In particular, the developed theoretical framework can be exploited to model exciton-exciton and exciton-phonon interactions governing the spatio-temporal dynamics of moir\'e excitons.

\section*{Acknowledgments}
We acknowledge funding from the Swedish Research Council (VR, project number 2018-00734), the European Unions Horizon 2020 research and innovation programme under grant agreement no. 881603 (Graphene Flagship), and the Knut and Alice Wallenberg Foundation (2014.0226).

\bibliographystyle{achemso}
\providecommand{\latin}[1]{#1}
\makeatletter
\providecommand{\doi}
  {\begingroup\let\do\@makeother\dospecials
  \catcode`\{=1 \catcode`\}=2 \doi@aux}
\providecommand{\doi@aux}[1]{\endgroup\texttt{#1}}
\makeatother
\providecommand*\mcitethebibliography{\thebibliography}
\csname @ifundefined\endcsname{endmcitethebibliography}
  {\let\endmcitethebibliography\endthebibliography}{}

\end{document}